\documentclass[sigconf,natbib=true]{acmart}
\AtBeginDocument{%
  \providecommand\BibTeX{{%
    \normalfont B\kern-0.5em{\scshape i\kern-0.25em b}\kern-0.8em\TeX}}}

\copyrightyear{2024}
\acmYear{2024}
\setcopyright{acmlicensed}\acmConference[SIGIR '24]{Proceedings of the 47th International ACM SIGIR Conference on Research and Development in Information Retrieval}{July 14--18, 2024}{Washington, DC, USA}
\acmBooktitle{Proceedings of the 47th International ACM SIGIR Conference on Research and Development in Information Retrieval (SIGIR '24), July 14--18, 2024, Washington, DC, USA}
\acmDOI{10.1145/3626772.3657683}
\acmISBN{979-8-4007-0431-4/24/07}

\usepackage{enumitem}
\usepackage{colortbl}  
\usepackage{xcolor}
\usepackage{url}
\usepackage{multirow}
\usepackage{subcaption}
\usepackage{arydshln}
\usepackage{tabularx}
\usepackage{enumitem}
\usepackage{array} 
\usepackage{booktabs}
\usepackage{pgfplots}  
\usepackage{graphicx}
\usepackage{float}
\usepackage{amsmath}
\usepackage{bm}

\begin{document}


\title{Large Language Models are Learnable Planners for Long-Term Recommendation}


\settopmatter{authorsperrow=4}

\author{Wentao Shi}
\affiliation{%
 \institution{University of Science and Technology of China}
 \city{Hefei}
 \country{China}}
\email{shiwentao123@mail.ustc.edu.cn}

\author{Xiangnan He}
\authornote{Corresponding author.}
\affiliation{%
 \institution{University of Science and Technology of China}
 \city{Hefei}
 \country{China}}
\email{xiangnanhe@gmail.com}

\author{Yang Zhang}
\affiliation{%
 \institution{University of Science and Technology of China}
 \city{Hefei}
 \country{China}}
\email{zy2015@mail.ustc.edu.cn}

\author{Chongming Gao}
\affiliation{%
 \institution{University of Science and Technology of China}
 \city{Hefei}
 \country{China}}
\email{chongming.gao@gmail.com}

\author{Xinyue Li}
\affiliation{%
 \institution{University of Science and Technology of China}
 \city{Hefei}
 \country{China}}
\email{lrel7@mail.ustc.edu.cn}

\author{Jizhi Zhang}
\affiliation{%
 \institution{University of Science and Technology of China}
 \city{Hefei}
 \country{China}}
\email{cdzhangjizhi@mail.ustc.edu.cn}

\author{Qifan Wang}
\affiliation{%
 \institution{Meta AI}
 \city{Menlo Park}
 \country{USA}}
\email{wqfcr@fb.com}

\author{Fuli Feng}
\authornotemark[1]
\affiliation{%
 \institution{University of Science and Technology of China}
 \city{Hefei}
 \country{China}}
\email{fulifeng93@gmail.com}

\renewcommand{\shortauthors}{Trovato and Tobin, et al.}
\newcommand{\zjz}[1]{{\color{blue}{#1}}}
\newcommand{\zy}[1]{{#1}}
\newcommand{\swt}[1]{\textcolor{yellow}{#1}}

\begin{abstract}
Planning for both immediate and long-term benefits becomes increasingly important in recommendation. Existing methods apply Reinforcement Learning (RL) to learn planning capacity by maximizing cumulative reward for long-term recommendation. However, the scarcity of recommendation data presents challenges such as instability and susceptibility to overfitting when training RL models from scratch, resulting in sub-optimal performance. 
In this light, we propose to leverage the remarkable planning capabilities over sparse data of Large Language Models (LLMs) for long-term recommendation. The key to achieving the target lies in formulating a guidance plan following principles of enhancing long-term engagement and grounding the plan to effective and executable actions in a personalized manner. To this end, we propose a Bi-level Learnable LLM Planner framework, which consists of a set of LLM instances and breaks down the learning process into macro-learning and micro-learning to learn macro-level guidance and micro-level personalized recommendation policies, respectively. Extensive experiments validate that the framework facilitates the planning ability of LLMs for long-term recommendation. Our code and data can be found at \url{https://github.com/jizhi-zhang/BiLLP}.
\end{abstract}




\maketitle
\section{Introduction}
Recommendation systems have gained widespread adoption in contemporary society to alleviate the overwhelming burden of information overload~\cite{DBLP:journals/corr/abs-2010-03240}. 
Traditionally, researchers primarily focused on optimizing users' immediate responses (\textit{e.g.} clicks) to maximize instant benefits~\cite{wu2017returning}.
However, such a greedy recommendation strategy tends to cater to users' immediate interests excessively, neglecting long-term engagement \cite{DBLP:conf/kdd/WangSXBSRCCC22} and even influencing the ecology negatively. For instance, some users will be confined within an echo chamber of preferred information and filter bubbles~\cite{DBLP:journals/tois/GaoWLCHLLZJ24}. Therefore, it is essential to investigate long-term recommendation.

To tackle this challenge, it is crucial to integrate planning capabilities into the recommendation decision-making process to develop policies that take into account not only immediate benefits but also long-term consequences.
Existing work primarily employs Reinforcement Learning~\cite{DBLP:conf/www/ZhengZZXY0L18, DBLP:journals/corr/abs-1810-12027, DBLP:conf/wsdm/ChenBCJBC19, DBLP:conf/sigir/WuX0ZZL22} to acquire planning capabilities implicitly through training models from scratch 
with the objective of maximizing cumulative rewards. 
However, these approaches are entirely data-driven, and their efficacy is significantly constrained by the quality and quantity of available data~\cite{DBLP:journals/corr/SchulmanWDRK17,DBLP:conf/sigir/GaoHCZLJW0023, DBLP:journals/tois/GaoWLCHLLZJ24}. Unfortunately, recommendation data is typically sparse and naturally long-tail distributed~\cite{DBLP:conf/www/0007WWC0023}. This poses a significant challenge for RL to acquire planning ability, particularly for sparse or long-tail items and users, resulting in sub-optimal performance.

LLMs have emerged with powerful planning capabilities through pre-training on massive and diverse textual data~\cite{RLHF,gpt4,llama2}.
Previous studies have demonstrated that LLMs 
can break down complex textual and agent tasks into subtasks and then 
execute them sequentially~\cite{DBLP:conf/corl/HuangXXCLFZTMCS22, DBLP:conf/icml/HuangAPM22, DBLP:journals/corr/abs-2211-09935, DBLP:journals/corr/abs-2302-01560}.
By conceptualizing multi-round recommendations as analogous to 
such complex tasks,
there is potential to harness the planning prowess of LLMs to devise a multi-round recommendation policy aimed at maximizing long-term engagement. 
Upon realization, benefitting from the inherent extensive world knowledge and robust reasoning capabilities in LLMs, it is anticipated to obtain superior planning capabilities even in scenarios with sparse recommendation data, especially for long-tail items.



\zy{To achieve the target, the key is to recall the task-solving principles to formulate a plan and make it effective and executable for individual users.} 
However, direct acquisition of such 
planning capabilities is non-trivial, due to the substantial 
scenario divergence between LLM pre-training and recommendation. 
In the realm of recommendation tasks, the LLM itself may not naturally exhibit an inherent understanding (or commonsense) of the principles that enhance long-term engagement. Additionally, when tailoring recommendations for individual users, a personalized and item-specific strategy becomes essential, far beyond the mere awareness of such guiding principles. It is necessary to inspire or teach the LLM to acquire the desired principles and make them personalized.

We propose a novel {\textbf{Bi}-level \textbf{L}earnable \textbf{L}LM \textbf{P}lanning} (BiLLP) framework for long-term recommendation. 
BiLLP breaks down the learning process into macro-learning and micro-learning through a hierarchical mechanism. 
Macro-learning, aiming at acquiring high-level guiding principles, includes a Planner and Reflector, both implemented as LLM instances. The Planner leverages memorized high-level experiences that imply guiding principles to formulate high-level plans for long-term goals, while the Reflector reflects on the finished trajectory to gather new experiences for updating the Planner. Micro-learning includes the LLM-based Actor-Critic component to acquire planning personalization. 
The Actor personalizes high-level plans into executable actions for users. The Critic functions similarly to the Reflector but operates on a more fine-grained level. It can promptly evaluate the long-term \textit{advantage} of an action given a state, facilitating the swift update of the Actor policy and mitigating high-variance issues in Q-values~\cite{DBLP:journals/corr/abs-1205-4839}.

\begin{figure*}[ht]
  \centering
  \includegraphics[width=0.8\textwidth]{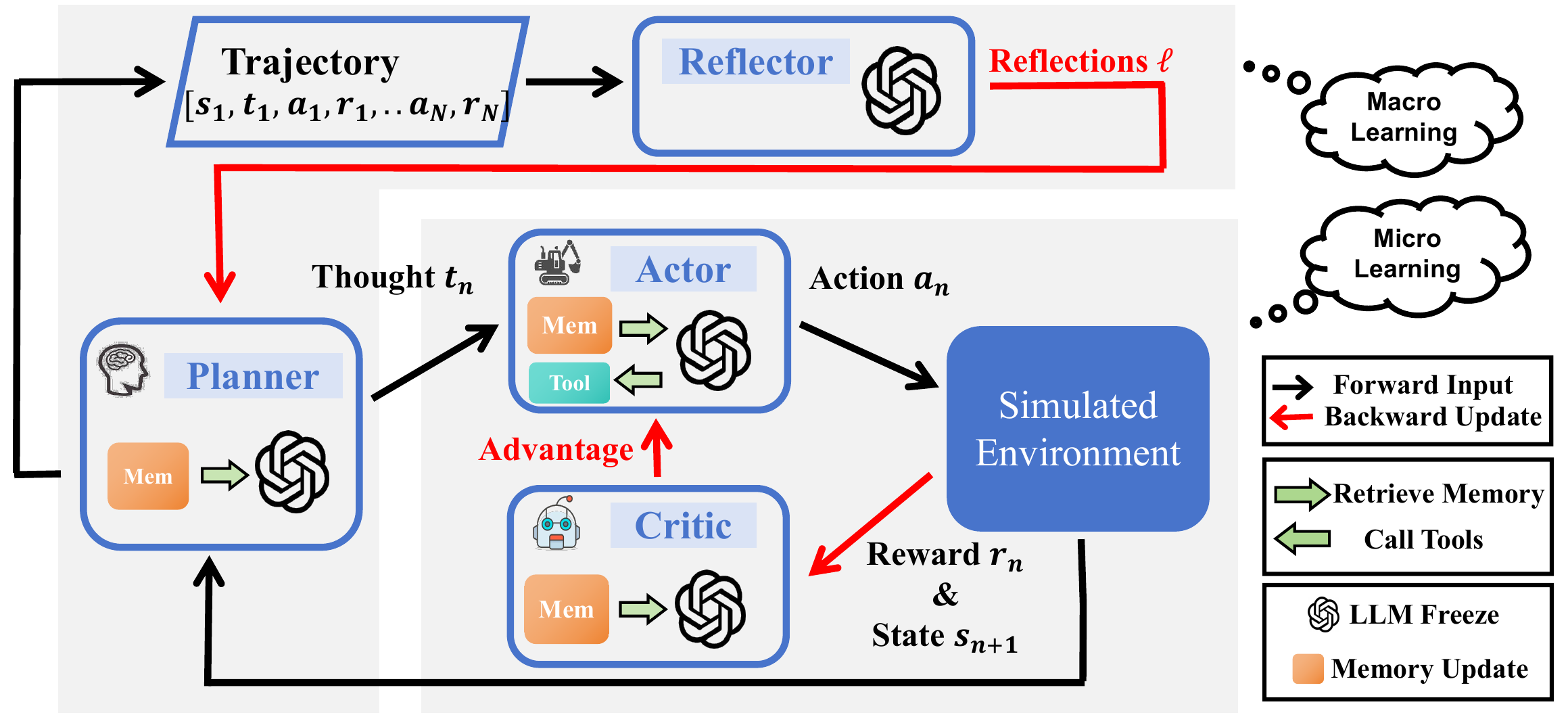}
  \vspace{-10pt}
  \caption{The overview of the proposed BiLLP framework. The black line indicates that the data serves as a prompt input for the subsequent module. The red line denotes that the data is utilized to update the memory of the subsequent module.}
  \vspace{-10pt}
  \label{fig:main_structure}
\end{figure*}

The main contributions of this work are summarized as follows:
\begin{itemize}[leftmargin=*]
    \item We introduce the idea of exploring the planning ability of LLMs with a bi-level planning scheme to enhance long-term engagement in recommendation.
    \item We propose a new BiLLP framework with four modules, which learns the planning ability at both macro and micro levels with low variance estimations of Q-values.
    \item We conduct extensive experiments, validating the capability of LLMs to plan for long-term recommendation and the superiority of the BiLLP framework. 
\end{itemize}










\section{Related Work}
\noindent $\bullet$ \textbf{Interactive Recommendation.}
Interactive recommendation is a typical setting to study long-term recommendation, where a model engages in online interactions with a user~\cite{gao2021advances,DBLP:conf/sigir/WuX0ZZL22}. 
In contrast to the static recommendation setting, where the focus is on identifying ``correct'' answers within a test set, interactive recommendation assesses the efficacy of results by accumulating rewards obtained throughout the interaction trajectories.
%
To improve the performance of interactive recommendation, extensive effort \cite{DBLP:journals/corr/Dulac-ArnoldESC15, DBLP:conf/ijcai/IeJWNAWCCB19, DBLP:conf/recsys/ZhaoXZDYT18} has been made to model the recommendation environment as a Markov decision process (MDP) and then utilize advanced RL algorithms to deliver the optimal policy \cite{DBLP:conf/www/ZhengZZXY0L18, DBLP:journals/corr/abs-1810-12027, DBLP:conf/wsdm/ChenBCJBC19, DBLP:conf/sigir/WuX0ZZL22}. CIRS~\cite{DBLP:journals/tois/GaoWLCHLLZJ24} learns a causal user model on historical data to capture the overexposure effect of items on user satisfaction, facilitating the planning of the RL policy. DORL~\cite{DBLP:conf/sigir/GaoHCZLJW0023} alleviates Matthew Effect of Offline RL to improve long-term engagement. However, these RL-based methods exhibit suboptimal learning efficiency and poor planning performance when confronted with sparse recommendation data.

\noindent $\bullet$ \textbf{LLM for Recommendation.}
LLM-based Recommendation paradigm has achieved remarkable advancements~\cite{survey_1, survey_2, survey_3} owing to the extraordinary abilities of LLMs such as GPT4~\cite{gpt4} and Llama2~\cite{llama2}. 
Distinguishing from existing LLM-based recommendation methods that are limited to directly using in-context learning~\cite{icl_1, icl_2, icl_3, icl_4, wei2023llmrec, xi2023towards} or tuning~\cite{it_1, it_2, it_3, it_4, wu2023exploring, kang2023llms} for immediate response in the recommendation, our proposed 
BiLLP delves deeply into how the powerful planning ability of LLMs can empower the long-term engagement of recommendation systems.
Some other approaches attempt to explore LLMs' planning capability in managing API tools~\cite{llm_agent_1, llm_agent_2, feng2023large} for recommendation, but they are also restricted to immediate responses and lack focus on users' long-term engagement.
In contrast to the previous approach of LLM-based recommendation, our proposed BiLLP deeply harnesses the planning capabilities of LLM and utilizes it to enhance long-term engagement for users which is particularly challenging to optimize in traditional recommendations.

\noindent $\bullet$ \textbf{LLM Planner.}
After undergoing pre-training and instruction tuning, the LLMs have attained extensive world knowledge and proficient planning capabilities. Recent work \cite{DBLP:conf/corl/HuangXXCLFZTMCS22, DBLP:conf/icml/HuangAPM22, DBLP:journals/corr/abs-2211-09935, DBLP:journals/corr/abs-2302-01560} exploit these powerful capabilities to generate better control plans for robots and agents. ReAct \cite{DBLP:conf/iclr/YaoZYDSN023} effectively integrates the action decision with planning and results in promising performance. SwiftSage \cite{DBLP:journals/corr/abs-2305-17390} integrates fast and slow thinking to solve complex tasks. However, these methods lack the ability to learn from past experiences, which allows for better task planning. To enable self-improvement without fine-tuning, Reflexion~\cite{shinn2023reflexion} verbally reflects on task feedback signals. ExpeL~\cite{DBLP:journals/corr/abs-2308-10144} utilize cross-task persistent memory to store insights and trajectories. AdaPlanner~\cite{DBLP:journals/corr/abs-2305-16653} can learn from past failure, past success, or both. In addition to these macro-level refinements, some work \cite{DBLP:journals/corr/abs-2306-07929, brooks2023large, DBLP:journals/corr/abs-2312-03290} integrates LLMs with RL algorithm to learn from the micro-level interaction experiences. However, they suffer from the issues of high variance estimations of Q-values, which could be alleviated by our proposed \textit{Critic module}.

\section{Preliminary}
\noindent $\bullet$ \textbf{Problem Definition.}
\label{subsection:problem_definition}
Following recent work on long-term recommendation~\cite{DBLP:journals/tois/GaoWLCHLLZJ24}, we adopt the interactive recommendation setting. The target is to learn a recommendation model that recommends items $i\in\mathcal{I}$ (\textit{i.e.}, makes an action\footnote{As an initial attempt, 
we constrain each action $a_n$ to recommend only one item.
} $a_n$) to a user $u\in\mathcal{U}$ at each step $n$ based on the current state\footnote{$s_1$ is the initial state before interacting with the model.} $s_n$.
%
As to applying LLMs for interactive recommendation, the recommendation process at each step $n$ involves two main operations: generating a problem-solving plan, referred to as thought $t_n$, and subsequently providing an item recommendation, denoted as action $a_n$. Based on this, the entire interaction episode can be denoted as \begin{equation}
    \mathcal{H}^{1\cdots N}=\{s_1,t_1,a_1,r_1,\cdots, s_N,t_N,a_N,r_N \},
\end{equation} 
and the trajectory $\mathcal{H}^{1\cdots n}, (1\leq n\leq N)$ can be thought of as a subsequence of an episode.

\noindent $\bullet$ \textbf{Simulated Environment.}
The interactive recommendation setting requires immediate user feedback for recommendation actions. Collecting online feedback from users can be financially burdensome, we thus follow~\cite{DBLP:journals/tois/GaoWLCHLLZJ24,DBLP:conf/sigir/GaoHCZLJW0023} and construct ``Simulated Environment'' with offline data for both model learning and testing. This environment can mimic users' behaviors in online scenarios, accept recommendations (\textit{i.e.} action $a$) from the model, and provide feedback (\textit{i.e.}, reward $r$) accordingly.

\section{Method}


In this section, we present the proposed BiLLP framework for improving long-term engagement in the interactive recommendation. As shown in Figure~\ref{fig:main_structure}, the recommendation process of our framework involves two main steps:
\begin{itemize}[leftmargin=*]
    \item The Planner generates problem-solving plans (\textit{i.e.}, thoughts $t$), where the recommendation task is broken into sequential step-by-step sub-plans, striking a harmonious balance between exploration and exploitation.
    \item The Actor recommends items (\textit{i.e.}, takes actions $a$) to the user by incorporating both macro-level sub-plans (thoughts) and micro-learning experiences.
\end{itemize}
%
To generate appropriate plans and personalized item recommendations, the key lies in teaching LLMs to learn from past interaction episodes. To enhance the learning process, the BiLLP framework employs a hierarchical mechanism (See Figure~\ref{fig:main_structure}):
\begin{itemize}[leftmargin=*]
    \item \textbf{Macro-learning} involves the Planner and Reflector to generate more appropriate plans, where the Reflector extracts high-level guiding principles from historical episodes and incorporates them into the input of Planner to enhance the quality of plans.
    \item \textbf{Micro-learning} involves the Actor and Critic to generate more personalized recommendations, where Critic assesses the user's current satisfaction level (action advantage value) and updates the policy of Actor to enhance personalized recommendations.
\end{itemize}

\subsection{Macro-Learning}

The macro-learning refers to a process in which the Reflector generates reflections based on historical episodes and subsequently updates them into the memory of the Planner. The Planner then retrieves the most relevant reflections from the memory and utilizes them as prompts to enhance the quality of plan generation. Next, we present the details of the Reflector and Planner, and the procedure of the micro-learning process.

\subsubsection{Reflector}\label{subsection:The Reflector Module}
The Reflector is designed to extract guiding principles from historical episode data. 
When a user ends his interaction with the model, 
we utilize this complete interaction episode
$\mathcal{H}_{c}^{1\cdots N}$ as input, and then generate reflections $\ell$ as follows:
\begin{equation}
    \ell_{c} = \textbf{Reflector}(\mathcal{H}_c^{1\cdots N}).
\end{equation}
We implement the Reflector as an LLM instance. Based on the predefined instruction prompt and few-shot examples $\mathcal{P}_{R}$, the reflection $\ell$ generation process can be formulated as:
\begin{equation}
    \ell_{c} = \text{LLM}(\mathcal{P}_{R}, \mathcal{H}_{c}^{1\cdots N}).
\end{equation}
%
%
%
The obtained reflections are then used to update the memory $\mathcal{M}_{P}$ in the Planner, denoted as $\ell_{c} \rightarrow \mathcal{M}_{P}$. 
To facilitate understanding, we provide an example of reflection in Table~\ref{tab:reflections}, which primarily covers two high-level aspects: \textit{analysis of withdrawal reasons} and \textit{prospective guidance}. Specifically, in the example, the users' disengagement is identified as stemming from the repetitive recommendation of identical items, and the guiding principle for future recommendations emphasizes prioritizing diversity. Both aspects do not involve specific items.

\begin{table}
\centering
  \caption{
        Example of reflections.
    }
    \vspace{-10pt}
 \label{tab:reflections}
    \begin{tabularx}{0.95\linewidth}{X}   
    \toprule
        \multicolumn{1}{c}{\textbf{Reflection Case 1}} \\
        \cdashline{1-1}[1pt/2.5pt]\noalign{\vskip 0.5ex} 
        The user became dissatisfied with the final recommendation, which was a repeat of a previously recommended game. This suggests that the user may have been looking for more variety in their recommendations. In the future, it would be beneficial to avoid repeating recommendations and instead focus on providing a diverse range of games across different genres.\\
    \bottomrule
    \end{tabularx}
    \vspace{-10pt}
\end{table}

\begin{table*}[t]
\centering
  \caption{
        Example of the input and output for the Planner module.
    }
 \label{tab:planner_module}
    \begin{tabularx}{0.95\textwidth}{lX}    
    \toprule
        \multicolumn{2}{c}{\textbf{Instruction Input}} \\
        \cdashline{1-2}[1pt/2.5pt]\noalign{\vskip 0.5ex} 
        \textbf{Instruction:}  & Solve a recommendation task with interleaving Thought, Action, and Observation steps. Thought can reason about the current situation and current user interest. Your goal is to meet the user's interest as much as possible and make recommendations to users as many times as possible. Note that if the user is not satisfied with your recommendations, he will quit and not accept new recommendations. You may take as many steps as necessary.\\
        & Here are some examples: \ \ \  <Few-shot Examples>  \ \ \  (END OF EXAMPLES) \\
        &Reflection: \{\textbf{Reflections $\ell
        _{\mathcal{M}_{P}}^{K}$}\} \\
        &\{\textbf{Historical interaction sequence $\mathcal{H}^{1\cdots n-1}$}\} \\  
        \midrule
        \multicolumn{2}{c}{\textbf{Output: Thought}}  \\
        \cdashline{1-2}[1pt/2.5pt]\noalign{\vskip 0.5ex}
        \textbf{Case 1:} & "The user seems to enjoy a mix of Action and Independent video games. They also seem to appreciate Adventure games. I would first recommend the user their favorite action games, and then recommend some other niche genre games that they like." \\
        \cdashline{1-2}[1pt/2.5pt]\noalign{\vskip 0.5ex}
        \textbf{Case 2:} & The user seems to be satisfied with the recommendations so far. Following the previous plan, I should recommend some other niche genre games that they like, such as RPG games..\\
    \bottomrule
    \end{tabularx}
    \vspace{-10pt}
\end{table*}

\subsubsection{Planner}\label{subsection:The Planner Module}
The Planner module is designed to generate forward-looking plans and decompose the high-level plan into sub-plans, indicated in outputted thoughts, where 
thoughts facilitate the Actor to execute actions. 
The Planner is implemented as a frozen LLM instance equipped with a memory library $\mathcal{M}_{P}$ storing past reflections for reference. 
At each step $n$ of a new episode, the Planner utilizes the historical trajectory $\mathcal{H}^{1\cdots n-1}$ and the current state $s_n$ from the environment as input to generate the thought $t_n$ with reflections obtained from the memory:
\begin{equation}
    t_n = \textbf{Planner}(\mathcal{H}^{1\cdots n-1}, s_n; \mathcal{M}_{P}),
\end{equation}
where the $\mathcal{M}_{P}$ is a set of past episode reflections. Formally, we have $\mathcal{M}_{P}=\{ \ell_m|m=1,2,...\}$.
When starting a new interaction process, meaning a new episode, multiple relevant reflections $\ell_{\mathcal{M}_{P}}^{K}$ are retrieved from the memory library $\mathcal{M}_{P}$ as guidance for generating new thoughts. For the following steps in the episode, we utilize the same reflections and other inputs to prompt LLM to generate thoughts. We next introduce these parts.

\noindent $\bullet$\textbf{Reflection retrieval}. To ensure that the retrieved reflections are helpful for planning. 
We select $K$ reflections with a minimal distance to this planning process. Taking the initial state $s_1$ to represent the process for distance computation, we have  
\begin{equation}
\begin{split}\label{eq:planner-retrieve}
     \ell_{\mathcal{M}_{P}}^{K} = \{ \ell | rank(d(\ell, s_1)) < K, \ell\in\mathcal{M}_{P}  \},
\end{split}
\end{equation}
where $d(\cdot, \cdot)$ is defined as the Euclidean distance between the two encoded texts
implemented by \textit{Facebook AI Similarity Search (FAISS)}~\cite{faiss}, a library that allows us to quickly search for similar documents.
$rank(\cdot)$ gets the rank of a value in ascending order.

\noindent $\bullet$\textbf{Thought generation}. We leverage the macro-level guidance from the memory to generate a thought. For each input $(\mathcal{H}^{1\cdots n-1}, s_n)$, we can sample a thought from LLM policy as follows: 
\begin{equation}
t_n \sim \text{LLM}(\mathcal{P}_{P}, \ell_{\mathcal{M}_{P}}^{K}, \mathcal{H}^{1\cdots n-1}, s_n).
\end{equation}
Here, $t_n$ is a sample from the Planner policy, the arguments in the function $\text{LLM}(\cdot)$ represent the prompt input to the LLM including task instruction of the Planner $\mathcal{P}_{P}$ (including few-shot examples), retrieved reflections $\ell_{\mathcal{M}_{P}}^{K}$, state $s_n$, and historical trajectory $\mathcal{H}^{1\cdots n-1}$ in the current episode.

Table~\ref{tab:planner_module} presents examples of our input prompt template and two representative thoughts, encapsulating common outputs of generated thoughts. In the input prompt, we integrate historical interaction sequences and reflections, prompting the LLM to generate appropriate thoughts for guiding subsequent actions. 
In these examples, we could find some interesting properties of the generated thoughts. 
In the case of thought example 1, we observe that LLM can analyze users' interests and decompose multiple rounds of recommendation tasks into distinct sub-plans. By leveraging its planning capabilities, the LLM can generate suggestions extending beyond immediate choices, considering their potential long-term impact on user satisfaction. This enables the method to take into account various factors to optimize user long-term engagement and satisfaction. In thought example 2, we note that LLMs can adhere to the previous plan, maintaining the consistency and continuity of the recommendation strategy.

\subsubsection{Update} The macro-learning involves updating the Planner during the training. After an episode is completed,
we update the Planner module by injecting the new reflections for this episode into its memory. The Planner memory update can be formulated as
\begin{equation}
    \mathcal{M}_{P} \leftarrow \ell_c,
\end{equation}
where $\ell_c$ denotes the reflections of the complete episode.



\subsection{Micro-Learning} 
The micro-learning refers to a process in which the Actor grounds the thoughts into executable actions to environments and the Critic provides evaluations of these actions.
By updating the policy of Actor based on the feedback of Critic and updating the policy of Critic based on the feedback of the environment,
Actor and Critic learn to provide personalized recommendations in specific situations. 
The learning mechanism is similar to the Planner-Reflector but operates in a more granular dimension, \textit{i.e.}, directly considering the recommendation of items. In essence, the micro-learning process bears analogies to the Advantage Actor-Critic (A2C) algorithm~\cite{DBLP:conf/icml/MnihBMGLHSK16}. In the following, we first introduce the details of the Actor and Critic modules and present the procedure of the micro-learning process.
\subsubsection{Actor} 
The Actor module aims to customize high-level plans into executable actions for each users. As illustrated in Figure~\ref{fig:main_structure}, similar to the Planner module, we implement it as an LLM instance equipped with a memory $\mathcal{M}_{A}$ storing micro-level experiences. Additionally, considering that some knowledge is valuable for personalization but challenging for LLMs to handle~\cite{DBLP:journals/corr/abs-2308-08434}, we add a tool library denoted as $Tl$ to access such knowledge. At each step $n$ of an episode, the actor utilizes the historical trajectory $\mathcal{H}^{1\cdots n-1}$, the current state $s_n$, and the corresponding thought $t_n$ from the Planner module as inputs to generate an executable action $a_{n}$ with knowledge obtained from the memory and the tool library as follows: 
$$a_n = \textbf{Actor}(\mathcal{H}^{1\cdots n-1}, s_n, t_n; \mathcal{M}_{A}, Tl). $$
Here, the memory $\mathcal{M}_{A}$ can be formulated as a set of micro-level experiences, where the $m$-th experience is a previous interaction record, including three factors: state $s_m$, action $a_{m}$, and corresponding value $v_{m}$. Formally, we have 
$\mathcal{M}_{A}=\{(s_m,a_m,v_m)|m=1,2,\dots\}$.

Upon receiving these inputs, each generation process comprises three operations: 1) retrieving valuable experiences from the memory $\mathcal{M}_{A}$, 2) utilizing the tools to gather valuable statistical information of the current state, and 3) integrating the results of the first two steps and other inputs to prompt LLM to generate an action. We next elaborate on these operations.

\noindent $\bullet$ \textbf{Retrieval.} Similar to the retrieval operation in the Planner module, we rely on the similarity between the experience and input to select valuable experiences from the memory. Specifically, we leverage the distance between the state of an experience and the input state to measure the similarity, and we select all experiences with distances smaller than a threshold $\tau_{A}$. The process can be formulated as follows:
\begin{equation}
 \Psi_{A}^{n} = \{(s_m, a_m, v_m)| d(s_m, s_{n})< \tau_{A}, s_m\in \mathcal{M}_{A}\},
\end{equation}
where $\Psi_{A}^{n}$ denotes the retrieved results, and $d(\cdot, \cdot)$ is the same to that in Equation~\eqref{eq:planner-retrieve}.




\noindent $\bullet$  \textbf{Tool analysis.} 
We utilize the tools in the tool library $Tl$ to analyze users' interaction history, extracting valuable information that is challenging for the LLM to handle. In this study, we primarily focus on leveraging the \textit{Category Analysis Tool}. At the $n$-th step, given the state $s_{n}$, the tool can identify a list of categories associated with each legal action and conduct statistical analysis on the user's viewing history. Formally,
\begin{equation}
    \mathcal{O}_n = Tl(s_n),
\end{equation}
where $\mathcal{O}_n$ denotes the tool output in text format. Notably, the methodology described here can be adapted and applied to various other tools.


\noindent $\bullet$  \textbf{Action generation.} 
We leverage both the guidance from the Planner module, micro-level knowledge obtained from the memory and tool to prompt the LLM of the Actor module to generate an action. For the input $(\mathcal{H}^{1\cdots n-1}, s_n, t_n)$ with $t_n$ representing the thought, once we obtain the corresponding retrieval results $\Psi_{A}^{n}$ and tool analysis result $O_{n}$, we can sample an action $a_n'$ from the LLM policy as follows:
\begin{equation}
    a_n' \sim \text{LLM}(\mathcal{P}_{A}, \Psi_{\mathcal{M}_{A}}^{n}, \mathcal{O}_n, \mathcal{H}^{1\cdots n-1}, s_n, t_n),
\end{equation}
where $\mathcal{P}_{A}$ represents the task instruction for the Actor. Note that the temperature coefficient of the LLM should be set to a non-zero value, ensuring the generation of non-deterministic results.

\textit{Item grounding.} The final action should be specific to an item within the candidate pool. Note that the LLM may generate items $a_n'$ that are not necessarily included in the pool. To address this issue, we adopt the grounding strategy from~\cite{DBLP:journals/corr/abs-2308-08434} to map $a_n'$ to an actual item with the highest similarity. Formally, the final action is obtained as follows:
\begin{align}
    a_n = \arg\min_{a\in \mathcal{I}} \text{sim}(\mathbf{e}_{a}, \mathbf{e}_{a_n'}),\quad
    \text{sim}(\mathbf{e}_{a},\mathbf{e}_{a_n'}) := \|\mathbf{e}_{a} - \mathbf{e}_{a_n'}\|,
\end{align}
where $\mathbf{e}_a$ represents the embedding of the action (item) $\alpha$ 
encoded by \textit{Llama2-7b}~\cite{llama2}, $\text{sim}(\cdot, \cdot)$ denotes the embedding similarity measured by the \textit{L2} distance, and $\|\cdot\|$ signifies the $L_2$ norm.




\subsubsection{Critic}\label{subsection:The Critic Module}

The Critic module is an LLM-based evaluator, providing evaluative feedback on the long-term goals for the actions generated by the Actor module to help update the policy of Actor. 
The Critic module also contains a memory $\mathcal{M}_{C}$ to store previous experiences.
Inspired by the A2C algorithm, we utilize the advantage value $v_n$ of action $a_n$ in the given state $s_n$ as the measurement. In particular, Critic takes the state $s_n$, action $a_n$, and the history trajectory $\mathcal{H}^{1\dots n-1}$ as inputs and then outputs the advantage $v_n$ with the experiences in $\mathcal{M}_{C}$ as references, which can be abstracted as follows:
\begin{equation}
    v_{n} = \textbf{Critic}(s_n, a_n; \mathcal{M}_{C}).
\end{equation}

To compute advantage values, similar to A2C, we first estimate the state-value function $V(s_{n})$, and then, based on it, we use the advantage function~\cite{DBLP:conf/icml/MnihBMGLHSK16} to determine the advantage value:

\noindent $\bullet$ \textbf{Estimating state-value.} 
The function $V(s_{n})$ provides an estimation of the value of being in state $s_n$ when following the Actor policy. We directly model the function with the LLM of the Critic module. In particular, we leverage in-context learning with few-shot examples and previous estimations in the memory $\mathcal{M}_C=\{(s_m, V(s_m))|m=1,2,\dots\}$ to predict the values of a given state $s_{n}$. Formally, we have:
\begin{equation}
    V(s_n) = \text{LLM}(\mathcal{P}_{C}, \Phi_{\mathcal{M}_{C}}^{n}, \mathcal{H}^{1\cdots n-1}, s_n),
    \label{eq:state_value_function}
\end{equation}
where $\mathcal{P}_{C} $ represents the used task prompt (including few-shot examples), and $\Phi_{\mathcal{M}_{C}}^{n}$ denotes the selected experiences from $\mathcal{M}_C$, which is obtained as follows:
\begin{equation}
    \Phi_{\mathcal{M}_{C}}^n = \{ (s_m, V(s_m))|d(s_m, s_n)<\tau_{C}, s_m\in \mathcal{M}_{C}   \},
\end{equation}
where $\tau_{C}$ is a threshold, and $d(\cdot, \cdot)$ denotes the same distance function in Equation~\ref{eq:planner-retrieve}.

\noindent $\bullet$ \textbf{Computing advantage value.} 
We next utilize the advantage function to determine the advantage value $v_n$ of action $a_n$ given the state $s_n$ at the $n$-th step. The advantage value is:
\begin{equation}
    v_n = \sigma(A(s_n, a_n)), \quad
   A(s_n, a_n) = r_{n} + \gamma * V(s_{n+1}) - V(s_n),
\end{equation}
where $A(\cdot,\cdot)$ is the commonly used advantage function, $r_{n}$ denotes the environmental reward at step $n$, and $s_{n+1}$ denotes the next-step state if taking action $a_{n}$ at state $s_n$. Regarding the function $\sigma$, we have $\sigma(x)=1$ if $x \geq 0$ else 0. Note that this approach mitigates the issue of high variance estimation of the Q-value in previous work~\cite{DBLP:conf/icml/MnihBMGLHSK16} (\textit{cf.} Section~\ref{subsection:critic_effect}).

\subsubsection{Update.}
The micro-learning involves updating both the Actor and the Critic during their iteration. At each step $n$, after obtaining the advantage value $v_n$ for the action $a_{n}$, we update the two modules by injecting the new experience into their memory.
\begin{itemize}[leftmargin=*]
    \item The Critic memory update can be formulated as
    \begin{equation}
        \mathcal{M}_{C} \leftarrow (s_n,  r_{n} + \gamma * V(s_{n+1})),
    \end{equation}
    where $r_{n} + \gamma * V(s_{n+1})$ can be considered as a more accurate estimation of   $V(s_n)$~\cite{DBLP:conf/icml/MnihBMGLHSK16}.
    \item The Actor memory update can be formulated as
    \begin{equation}
        \mathcal{M}_{A} \leftarrow (s_n, v_{n}).
    \end{equation}
\end{itemize}
The updated memory incorporates new experiences, helping enhance the next step of processing.

\subsection{Discussion}
\begin{figure}[t]
  \centering
  \includegraphics[width=\linewidth]{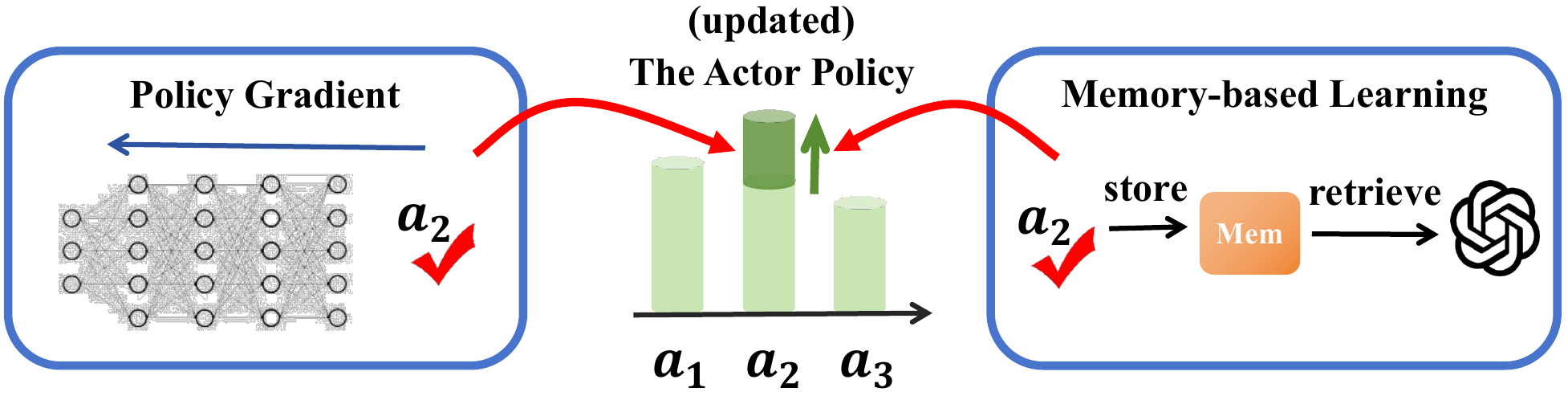}
  \caption{The memory-based learning methods and policy gradient based methods have a comparable impact on the Actor policy. }
  \label{fig:analogy}
  \vspace{-10pt}
\end{figure}

Next, we compare the policy update of our BiLLP framework with traditional gradient-based policy updates to illustrate why our approach, based on in-context learning, can learn planning without the need for gradient updates. As shown in Figure~\ref{fig:analogy}, for traditional methods, when a favorable action is identified for a state (possibly determined based on Q-values in the REINFORCE algorithm \cite{DBLP:journals/corr/MnihKSGAWR13} or the Advantage Function in A2C \cite{DBLP:conf/icml/MnihBMGLHSK16}), the purpose of the gradient update is to adjust the policy to increase the probability of sampling that specific action for the given state. In contrast, for our method, although no gradient updates are performed, the specific state and action are recorded in external memory. When encountering a similar state again, the retrieving probability of that specific state and action from the memory will increase, which would further enhance the probability of executing that specific action in that state. This achieves a similar effect to gradient updates. This is the underlying learning principle for our BiLLP framework.

\section{Experiments}

In this section, we evaluate the proposed BiLLP framework in the interactive recommendation settings. Our experiments aim to address the following questions:
\begin{itemize}[leftmargin=*]
    \item \textbf{RQ1:} How does BiLLP perform compared to state-of-the-art RL-based methods and other LLM frameworks in the interactive recommendation setting?
    \item \textbf{RQ2:} To what extent can macro-learning and micro-learning mechanisms improve the LLMs' planning ability?
    \item \textbf{RQ3:} Can the proposed Critic module effectively estimate the state-value function to facilitate the update of the Actor module?
    \item \textbf{RQ4:} Whether the proposed BiLLP framework is robust to different recommendation environments and base LLM models?
\end{itemize}

\subsection{Experiments Setup}
We introduce the experimental settings with regard to simulated experiments and baselines, which are implemented based on the EasyRL4Rec library\footnote{\url{https://github.com/chongminggao/easyrl4rec}} \cite{yu2024easyrl4rec}.

\subsubsection{Recommendation Experiments}\label{subsection:recommendation_experiments}
In the interactive recommendation setting, we are interested in examining the potential of models to mitigate the issue of filter bubbles and maximize users' long-term engagement. Conducting direct online experiments for model learning and testing can be prohibitively costly. As a result, following \cite{DBLP:conf/sigir/GaoHCZLJW0023}, we resort to creating simulated interactive environments using high-quality logs.
\begin{itemize}[leftmargin=*]
    \item \textbf{Steam}~\cite{DBLP:conf/icdm/KangM18} contains reviews and game information. The dataset compiles titles and genres of games. we consider users who engage in gameplay for a duration exceeding 3 hours to have a rating of 5, while others are assigned a rating of 2. We filter out users and items that interact less than 5 times in the log. 
    \item \textbf{Amazon-Book}~\cite{DBLP:conf/emnlp/NiLM19} refers to a book recommendation dataset, the ``book'' subset of the famous Amazon Product Review dataset\footnote{\url{https://cseweb.ucsd.edu/~jmcauley/datasets/amazon_v2/}}. This dataset compiles titles and genres of books from Amazon, collected between 1996 and 2018, with review scores ranging from 1 to 5. We filter out users and items that interact less than 90 times in the log. 
\end{itemize}
To better reflect the issue of filter bubbles and simulate real-world recommendation scenarios, we follow \cite{DBLP:journals/tois/GaoWLCHLLZJ24, DBLP:conf/sigir/GaoHCZLJW0023, DBLP:conf/cikm/XuTFJHZ22} to introduce a quit mechanism. The interaction will terminate if any of the following conditions are met:
\begin{itemize}[leftmargin=*]
    \item The similarity between a recommended item and the items in the recent recommendation list (with a window size of $W$) is below a predefined threshold $\beta$.
    \item The online reward $r$ of a recommended item is less than 2.
\end{itemize}
In this sense, a model that effectively captures users' interests and mitigates the risk of continuously recommending similar items that reinforce the filter bubble phenomenon is crucial for achieving a longer interaction trajectory and maximizing cumulative rewards. To estimate the online reward, we first split the dataset evenly into training and test sets in chronological order. For each set $\mathcal{D}\in \{ \mathcal{D}_{train}, \mathcal{D}_{test}\}$, we utilize the DeepFM model \cite{DBLP:conf/ijcai/GuoTYLH17} to fit the data and obtain vector representations for users $\mathbf{e}_u^{\mathcal{D}}$ and items $\mathbf{e}_i^{\mathcal{D}}$. Then we can calculate the online rewards:
\begin{equation}
    r_{u,i}^{\mathcal{D}} = \text{DeepFM}(\mathbf{e}_u^{\mathcal{D}}, \mathbf{e}_i^{\mathcal{D}}),\ \ \  u\in\mathcal{U}, i\in\mathcal{I},
\end{equation}
and the similarity between the two items:
\begin{equation}
    \text{sim}(\mathbf{e}_i^{\mathcal{D}}, \mathbf{e}_j^{\mathcal{D}}) = |\mathbf{e}_i^{\mathcal{D}} - \mathbf{e}_j^{\mathcal{D}}|_2, \ \ \ i,j\in \mathcal{I},
\end{equation}
It is noteworthy that we have established separate training and test environments for each dataset in order to simulate real-world scenarios where the user interests may have evolved during online training and model deployment. For now, the simulated environments can play the same role as the online users. Therefore, we can train the model on the training simulated experiments and evaluate the model on the test simulated experiments as the process shown in Figure~\ref{fig:main_structure}. The statistics of the datasets are illustrated in Table~\ref{tab:dataset}.

 \begin{table*}[ht]
\centering
  \caption{Average results of all methods in two environments (Bold: Best, Underline: Runner-up).}
  \vspace{-10pt}
  \label{tab:main_result}
  \begin{tabular}{lcccccc}    
    \toprule
    \multirow{2}*{Methods} & \multicolumn{3}{c}{Steam} & \multicolumn{3}{c}{Amazon} \\
    \cmidrule(lr{1em}){2-4} \cmidrule(lr{1em}){5-7}
    & Len & $\mathrm{R_{each}}$ & $\mathrm{R_{traj}}$ & Len & $\mathrm{R_{each}}$ & $\mathrm{R_{traj}}$ \\ 
    \midrule
    SQN & 2.183  $\pm$  0.177 & 3.130 $\pm$ 0.050 & 6.837 $\pm$ 0.517 & 4.773 $\pm$ 0.059 & 4.303 $\pm$ 0.017 & 20.570 $\pm$ 0.245 \\
    CRR & 4.407 $\pm$ 0.088 & 3.263 $\pm$ 0.427 & 14.377 $\pm$ 1.658 & 3.923 $\pm$ 0.162 & 4.537 $\pm$ 0.103 & 17.833 $\pm$ 1.129 \\
    BCQ & 4.720 $\pm$ 0.343 & 3.997 $\pm$ 0.068 & 18.873 $\pm$ 1.092 & 4.847 $\pm$ 0.721 & 4.367 $\pm$ 0.053 & 21.150 $\pm$ 2.893 \\
    CQL & 5.853 $\pm$ 0.232 & 3.743 $\pm$ 0.147 & 21.907 $\pm$ 0.299 & 2.280 $\pm$ 0.185 & 4.497 $\pm$ 0.039 & 10.263 $\pm$ 0.882 \\
    DQN & 4.543 $\pm$ 0.693 & 4.500 $\pm$ 0.069 & 20.523 $\pm$ 3.618 & 4.647 $\pm$ 0.498 & 4.290 $\pm$ 0.083 & 19.923 $\pm$ 1.909 \\
    A2C & 9.647 $\pm$ 0.848 & 4.367 $\pm$ 0.069 & 42.180 $\pm$ 3.937 & 7.873 $\pm$ 0.310 & 4.497 $\pm$ 0.026 & 35.437 $\pm$ 1.453 \\
    DORL & 9.467 $\pm$ 0.862 & 4.033 $\pm$ 0.098 & 38.300 $\pm$ 4.173 & 7.507 $\pm$ 0.174 & 4.510 $\pm$ 0.014 & 33.887 $\pm$ 0.655 \\
    \midrule
    ActOnly & 5.567 $\pm$ 0.160 & \underline{4.537 $\pm$ 0.021} & 25.250 $\pm$ 0.637 & 6.383 $\pm$ 0.176 & 4.490 $\pm$ 0.008 & 28.660 $\pm$ 0.761 \\
  ReAct & 11.630 $\pm$ 0.741 & \textbf{4.559 $\pm$ 0.047} & 52.990 $\pm$ 2.925 & 7.733 $\pm$ 0.450 & \underline{4.603 $\pm$ 0.033} & 35.603 $\pm$ 1.806 \\
    Reflexion & \underline{12.690 $\pm$ 1.976} & 4.523 $\pm$ 0.026 & \underline{57.423 $\pm$ 8.734} & \underline{8.700 $\pm$ 0.535} & \textbf{4.670 $\pm$ 0.073} & \underline{40.670 $\pm$ 2.954} \\
    BiLLP & \textbf{15.367 $\pm$ 0.119} & 4.503 $\pm$ 0.069 & \textbf{69.193 $\pm$ 1.590} & \textbf{9.413 $\pm$ 0.190} & 4.507 $\pm$ 0.012 & \textbf{42.443$\pm$ 0.817} \\
    \bottomrule
  \end{tabular}
\end{table*}

\begin{table}
  \centering
  \caption{Statistics of experiment datasets.}
  \vspace{-5pt}
  \label{tab:dataset}
  \begin{tabular}{lcccc}    
    \toprule
    Datasets & \#Users & \#Items  &\#Train &\#Test \\
    \midrule
    Steam & 6,012 & 190,365 & 1,654,303 & 958,452 \\
    Amazon &3,109 & 13,864 & 339,701 & 137,948 \\
    \bottomrule
\end{tabular}
\vspace{-10pt}
\end{table}

\subsubsection{Evaluation Metrics}
In this paper, we utilize three metrics: the \textbf{trajectory length} (Len), the average \textbf{single-round reward} ($\mathrm{R_{each}}$), and the \textbf{cumulative reward} of the whole trajectory ($\mathrm{R_{traj}}$) to evaluate the model performance in the interactive recommendation setting. Longer trajectory lengths and higher cumulative rewards demonstrate the model's ability to maximize long-term engagement. However, it is important to note that a higher average reward is not necessarily better. Excessively high average rewards may indicate a model's overemphasis on immediate responses.

\subsubsection{Baselines}
To comprehensively and fairly evaluate the superiority of our proposed BiLLP, we choose some representative RL-based methods and LLM-based methods as baselines.
For the RL-based methods, we choose seven representative baselines including the State-Of-The-Art (SOTA) method for long-term engagement optimization to mitigate filter bubble problems:
\begin{itemize}[leftmargin=*]
    \item \textbf{DQN}, or Deep Q-Networks \cite{DBLP:journals/corr/MnihKSGAWR13}, is a deep reinforcement learning algorithm that combines deep neural networks with the Q-learning algorithm.
    \item \textbf{SQN}, or Self-Supervised Q-learning \cite{DBLP:conf/sigir/XinKAJ20}, consists of two output layers, namely the cross-entropy loss head and the RL head. The RL head is utilized to generate the final recommendations.
    \item \textbf{BCQ}, or Batch-Constrained deep Q-learning \cite{DBLP:conf/icml/FujimotoMP19}, a modified version of conventional deep Q-learning designed for batch reinforcement learning. It utilizes the discrete-action variant \cite{DBLP:journals/corr/abs-1910-01708}, which focuses on discarding uncertain data and updating the policy solely based on high-confidence data.
    \item \textbf{CQL}, or Conservative Q-Learning \cite{DBLP:conf/nips/KumarZTL20}, is a model-free RL method that adds a Q-value regularizer on top of an actor-critic policy.
    \item \textbf{CRR}, or Critic Regularized Regression \cite{DBLP:conf/nips/0001NZMSRSSGHF20}, is a model-free RL method that learns the policy by avoiding OOD actions.
    \item \textbf{A2C}, or Advantage Actor-Critic \cite{DBLP:conf/icml/MnihBMGLHSK16}, improves the Actor-Critic algorithm and stabilizes learning by using the Advantage function as Critic instead of the Action value function.
    \item \textbf{DORL}, or Debiased model-based Offline RL \cite{DBLP:conf/sigir/GaoHCZLJW0023}, add a penalty term to relax the pessimism on states with high entropy to alleviate the Matthew effect in offline RL-based recommendation. This is the SOTA method of maximizing users' long-term engagement to alleviate filter bubble issues.
\end{itemize}
Ensuring fair comparison, we also implement three LLM-based baselines utilizing the same LLM backbone as BiLLP:
\begin{itemize}[leftmargin=*]
    \item \textbf{ActOnly}, a baseline that recommends items to users according to instruction prompts without thought and planning.
    \item \textbf{ReAct}, \cite{DBLP:conf/iclr/YaoZYDSN023} utilizes LLMs to generate both reasoning traces and task-specific actions in an interleaved manner, allowing for greater synergy between the reasoning and acting.
    \item \textbf{Reflexion}, \cite{shinn2023reflexion} verbally reflects on task feedback signals, then maintains their own reflective text in an episodic memory buffer to induce better decision-making in subsequent trials.
\end{itemize}

\subsubsection{Implementation Details}\label{subsection:hyperparameter}
For a fair comparison, all RL-based methods are trained with 100,000 episode data, and all LLM-based methods are trained with 100 episode data. For model-based RL methods DORL, we use the same DeepFM model as~\cite{DBLP:conf/sigir/GaoHCZLJW0023}. For the Reflection and BiLLP methods, we set the number of most similar reflections $K=2$. For the BiLLP method, we set the similarity threshold $\tau_{A}=0.01$ and $\tau_{C}=0.1$. The discount factor $\gamma$ is set to 0.5.
All methods in two environments are evaluated with the quit parameters: $W=4$, $\beta_{Steam}=50$, and $\beta_{Amazon}=15$. The maximum round is
set to 100. 
For all the RL-based methods, we leverage DeepFM~\cite{DBLP:conf/ijcai/GuoTYLH17} as the backbone following~\cite{DBLP:conf/sigir/GaoHCZLJW0023}, and for all the LLM-based methods, we utilize the ``gpt-3.5-turbo-16k'' provided by OpenAI as the LLM backbone for its strong long context modeling ability. And the temperature is set to 0.5 for all experiments.

\subsection{Main Results Comparison (RQ1)}
After training, we evaluate all methods with 100 episodes (\textit{i.e.}, interaction trajectories) in two interactive environments. The results are shown in Table~\ref{tab:main_result}, where each result in the table is averaged over three random experiments with distinct seeds for robustness and reliability. From the results, we observe that:
\begin{itemize}[leftmargin=*]
    \item BiLLP consistently achieves the best long-term performance (Len and $\mathrm{R_{traj}}$) over RL-based methods and LLM-based baselines across two datasets. This demonstrates the effectiveness of our proposed framework and its ability to stimulate and adapt the long-term planning capacity of LLMs. For single-round reward $\mathrm{R_{each}}$, BiLLP obtains a relatively higher score, which indicates it successfully captures user interests while avoiding excessive emphasis on immediate responses.
    \item The ActOnly method, which only utilizes LLMs to generate actions (recommendations), exhibits inferior performance compared to certain RL-based methods and LLM-based methods that incorporate planning. Drawing upon this, we can infer that an explicit thinking and planning process is crucial for enhancing the planning capabilities of LLMs.
    \item The ReAct method, which integrates the thinking process and action process, still performs worse than Reflexion and BiLLP. This underscores the significance of self-improvement in LLMs, in order to improve their planning abilities for long-term recommendation tasks.
\end{itemize}

\begin{figure}
    \begin{minipage}[t]{0.48\linewidth}
        \centering
        \includegraphics[width=\linewidth]{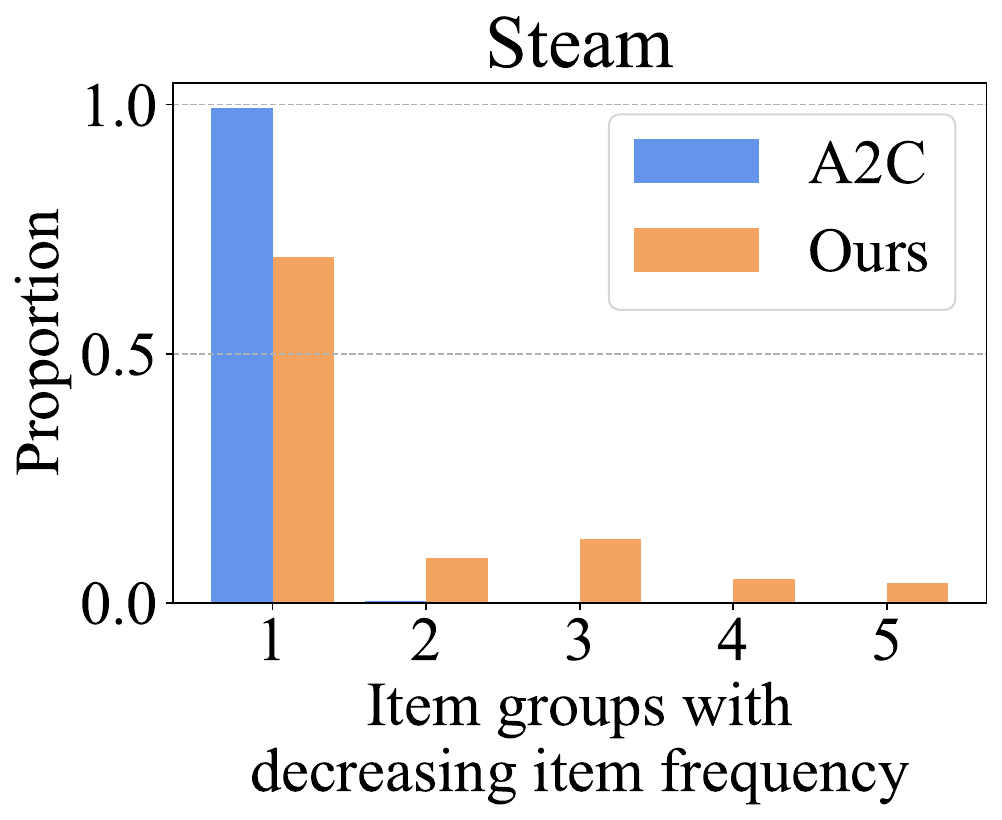}
        \label{fig:enter-label}
    \end{minipage}
    \begin{minipage}[t]{0.48\linewidth}
        \centering
        \includegraphics[width=\linewidth]{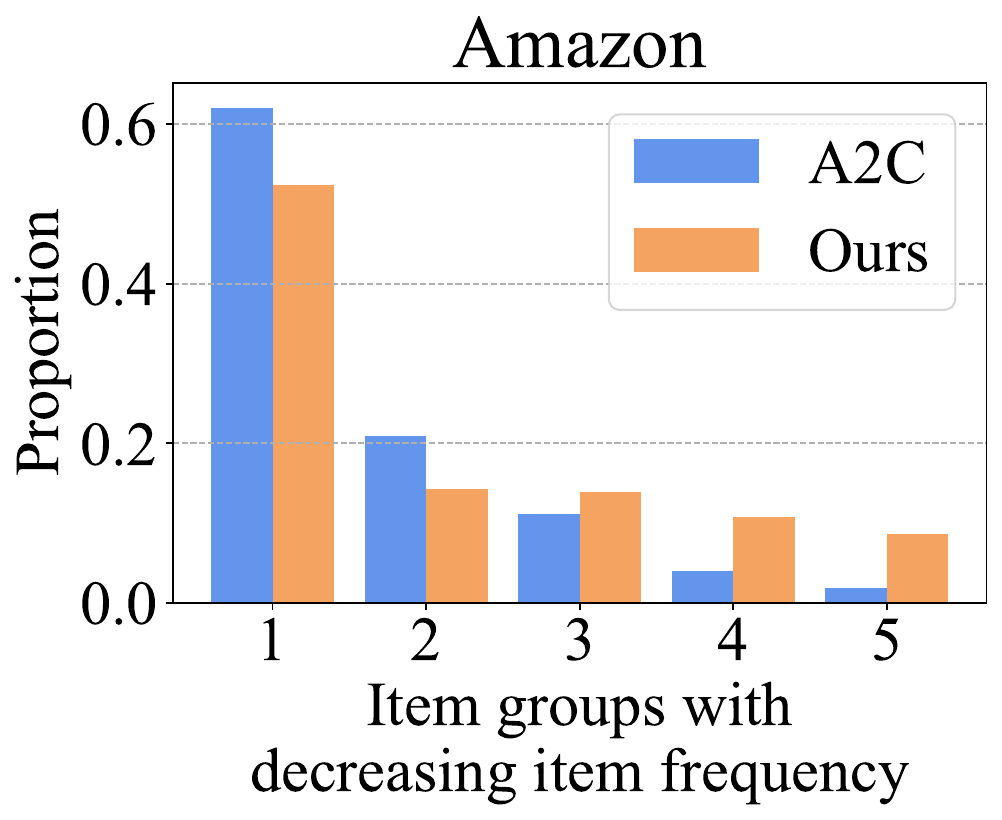}
        \label{fig:enter-label}
    \end{minipage}
    \vspace{-10pt}
    \caption{The frequency distribution of items recommended by our method and A2C in the two environments.}
    \label{fig:popularity_analysis}
    \vspace{-10pt}
\end{figure}
In addition to the overall performance comparison, we conduct an in-depth analysis of the recommended items for RL-based method A2C and LLM-based method BiLLP. We first calculate the items' popularity (occurrence frequencies) both in the training set and test set. Subsequently, we evenly divide the items into five groups with decreasing popularity: 1, 2, 3, 4, 5. We analyze the proportion of items belonging to each group among the recommended items generated by A2C and BiLLP, where the results are shown in Figure~\ref{fig:popularity_analysis}. From the figure, we observe that:
\begin{itemize}[leftmargin=*]
    \item RL-based method A2C tends to overfit on popularity items and lack planning capabilities on long-tail items.
    \item In contrast, BiLLP exhibits better planning capabilities on long-tail items, which could effectively alleviate the issue of filter bubbles and maximize long-term engagement.
\end{itemize}

\subsection{Ablation Study (RQ2)}
In this subsection, we conduct ablation studies to evaluate the effect of the two learning mechanisms. Concretely, \textbf{w/o Macro} refers to a variant of the BiLLP framework that does not use the reflective text to enhance its Planner module, and \textbf{w/o Micro} refers to a variant that does not use the micro-learning experience to enhance its Actor module. From Table~\ref{tab:ablation_study}, we can observe that:
\begin{itemize}[leftmargin=*]
    \item The absence of either of the two learning mechanisms would result in a decline in performance, thereby indicating that both learning mechanisms have contributed to the enhancement of long-term engagement.
    \item Based on the experimental details presented in Section~\ref{subsection:hyperparameter}, the aforementioned improvements are achieved using only 100 episodes of data for both learning mechanisms. This suggests the high efficiency of in-context learning compared to fine-tuning and training from scratch.
\end{itemize}

\begin{table}
  \centering
  \caption{Average results of all methods in the two environments (Bold: Best).}
  \vspace{-10pt}
  \label{tab:ablation_study}
  \begin{tabular}{lccc}    
    \toprule
    \multirow{2}*{Methods} & \multicolumn{3}{c}{Steam}\\
    \cmidrule(lr{1em}){2-4}
    & Len & $\mathrm{R_{each}}$ & $\mathrm{R_{traj}}$  \\ 
    \midrule
    w/o Macro & 14.363 $\pm$ 0.467  & 4.523 $\pm$ 0.012  & 64.960 $\pm$ 2.011 \\
    w/o Micro  & 14.270 $\pm$ 0.190  & \textbf{4.535 $\pm$ 0.005}  & 64.720 $\pm$ 0.920  \\
    BiLLP  & \textbf{15.367 $\pm$ 0.119} & 4.503 $\pm$ 0.069 & \textbf{69.193 $\pm$ 1.590}    \\
    \midrule
    \multirow{2}*{Methods} & \multicolumn{3}{c}{Amazon} \\
    \cmidrule(lr{1em}){2-4}
    & Len & $\mathrm{R_{each}}$ & $\mathrm{R_{traj}}$  \\ 
    \midrule
    w/o Macro  & 8.947 $\pm$ 0.480  & 4.530 $\pm$ 0.057 & 40.547 $\pm$ 2.622\\
    w/o Micro  & 8.800 $\pm$ 0.432  & \textbf{4.707 $\pm$ 0.026}   & 41.420 $\pm$ 2.003   \\
    BiLLP  & \textbf{9.413 $\pm$ 0.190} & 4.507 $\pm$ 0.012 & \textbf{42.443 $\pm$ 0.817}    \\
    \bottomrule
  \end{tabular}
  \vspace{-10pt}
\end{table}

\subsection{Effects of Critic Module (RQ3)}\label{subsection:critic_effect}
\begin{figure}[t]
  \centering
  \includegraphics[width=\linewidth]{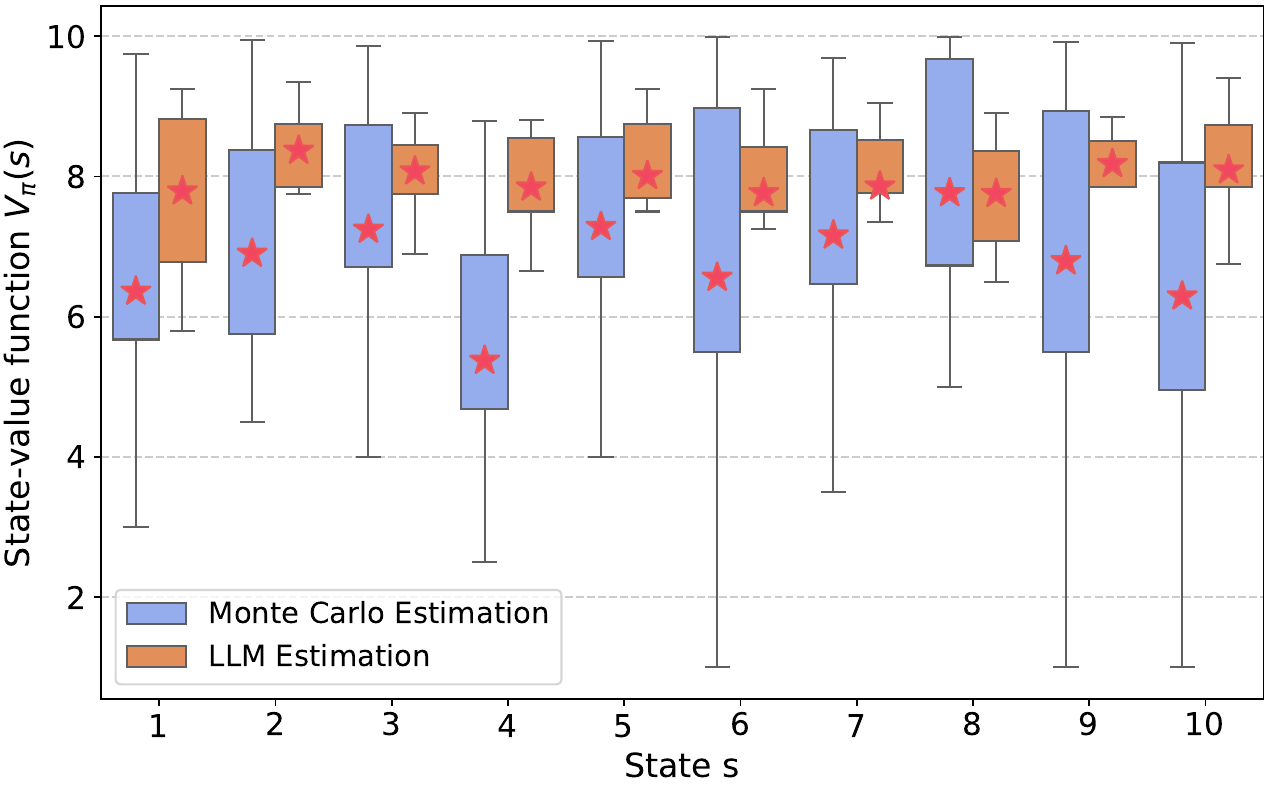}
  \vspace{-10pt}
  \caption{The memory-based in-context learning methods and policy gradient-based methods have a comparable impact on the Actor policy. }
  \label{fig:critic_estimation}
  \vspace{-10pt}
\end{figure}
In this subsection, our objective is to demonstrate the effectiveness of the Critic module in estimating the state-value function, denoted as $V_\pi(s)$, which is crucial for facilitating the update process of the Actor module. 
The state-value function $V_\pi(s)$ gives the expected cumulative discounted reward if we start from state $s$ and act according to the policy. 

To get an accurate and unbiased estimation of $V_\pi(s)$, for a specific state $s$, we sample 1000 complete trajectories according to the Actor policy and calculate the cumulative discounted reward for each trajectory. Figure~\ref{fig:critic_estimation} illustrates the distribution of these reward samples, along with their mean value. The mean value serves as an accurate and unbiased estimation of the state-value function $V_\pi(s)$. It is worth noting that prior studies \cite{brooks2023large, DBLP:journals/corr/abs-2306-07929}, have utilized a single trajectory's cumulative discounted reward to estimate either the state-value function $V_\pi(s)$ or the state-action value function $Q_\pi(s,a)$, which suffers from the issue of high variance estimation.

In contrast to these approaches, we leverage the Critic module to estimate the state-value function $V_\pi(s)$.
To evaluate our estimation, we repeat the estimation 100 times. The resulting estimations, as well as their distribution and mean value, are also depicted in Figure~\ref{fig:critic_estimation}. Based on the analysis of ten different states, it can be inferred that the utilization of the Critic module effectively mitigates estimation variance, despite the presence of a small bias in the estimation.

\subsection{Robustness of the Framework (RQ4)}


\subsubsection{Results with Different Environments}
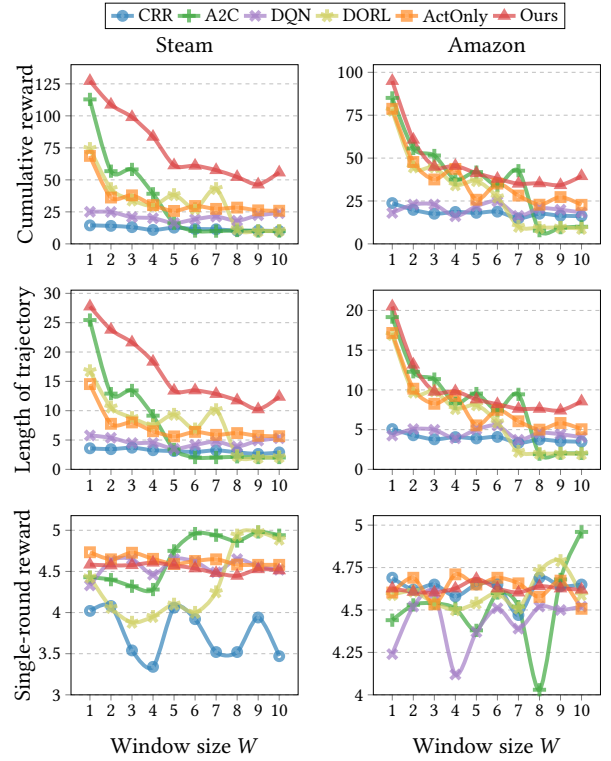
\begin{figure}[t]
     
     \begin{tikzpicture}[scale=0.5]
        \definecolor{c1}{RGB}{51,131,186}
        \definecolor{c2}{RGB}{63,168,63}
        \definecolor{c3}{RGB}{164,126,198}
        \definecolor{c4}{RGB}{208,208,100}
        \definecolor{c5}{RGB}{255,138,36}
        \definecolor{c6}{RGB}{219,69,70}
         \begin{axis}[
             name = ax1,
             ylabel = \Huge{Cumulative reward},
             every axis plot/.append style={line width = 3pt},
             tick align=outside,
             ymin = 0,
             ytick distance = 25,
             xtick distance = 1,
             xtick pos = left,
             ytick pos = left,
             ymajorgrids,
             tick label style = {font = \huge},
             grid style = dashed,
             width=3in,
             height=2.5in,
             legend style = {at = {(0.15,1.2)},anchor = south west, legend columns = -1, font = \huge},
             title = \Huge{Steam},
             ]
        \addplot[smooth, mark=o, mark size = 3pt, c1, draw opacity = 0.7] plot coordinates {
            (1, 14.44)
            (2, 14.05)
            (3, 13.07)
            (4, 10.89)
            (5, 12.67)
            (6, 11.63)
            (7, 11.42)
            (8, 10.25)
            (9, 10.35)
            (10, 10.04)
        };
        \addlegendentry{CRR}

         \addplot[smooth, mark=+, mark size = 5pt, c2, draw opacity = 0.7] plot coordinates {
            (1, 112.86)
            (2, 56.79)
            (3, 58.16)
            (4, 39.14)
            (5, 16.11)
            (6, 10.02)
            (7, 9.97)
            (8, 10.43)
            (9, 10.06)
            (10, 9.98)
        };
        \addlegendentry{A2C}       

        \addplot[smooth, mark=x, mark size = 5pt, c3, draw opacity = 0.7] plot coordinates {
            (1, 25.03)
            (2, 24.92)
            (3, 20.88)
            (4, 20.02)
            (5, 16.32)
            (6, 19.22)
            (7, 21.29)
            (8, 18.07)
            (9, 22.35)
            (10, 23.95)
        };
        \addlegendentry{DQN}

        \addplot[smooth, mark=star, mark size = 5pt, c4, draw opacity = 0.7] plot coordinates {
            (1, 74.35)
            (2, 43.02)
            (3, 34.11)
            (4, 29.76)
            (5, 38.33)
            (6, 27.4)
            (7, 43.08)
            (8, 11.47)
            (9, 10.01)
            (10, 10.31)
        };
        \addlegendentry{DORL}
        
        \addplot[smooth, mark=square, mark size = 3pt, c5, draw opacity = 0.7] plot coordinates {
            (1, 68.733)
            (2, 36.2)
            (3, 37.910)
            (4, 30.253)
            (5, 25.723)
            (6, 29.593)
            (7, 27.330)
            (8, 28.333)
            (9, 26.180)
            (10, 25.880)
        };
        \addlegendentry{ActOnly}

         \addplot[smooth, mark=triangle, mark size = 3pt, c6, draw opacity = 0.7] plot coordinates {
            (1, 127.250)
            (2, 108.757)
            (3, 99.027)
            (4, 83.640)
            (5, 61.357)
            (6, 61.053)
            (7, 57.613)
            (8, 52.263)
            (9, 46.397)
            (10, 55.780)
        };       
        \addlegendentry{Ours}
     \end{axis}
     
    \begin{axis}[
        name = ax2,
        at = {(ax1.south east)},
        xshift = 2cm,
        tick label style = {font = \huge},
        every axis plot/.append style={line width = 3pt},
        tick align=outside,
        ymin = 0,
        ytick distance = 25,
        xtick distance = 1,
        xtick pos = left,
        ytick pos = left,
        ymajorgrids,
        grid style = dashed,
        width=3in,
        height=2.5in,
        title = \Huge{Amazon}
        ]
        
        \addplot[smooth, mark=o, mark size = 3pt, c1, draw opacity = 0.7] plot coordinates {
            (1, 23.79)
            (2, 19.82)
            (3, 17.57)
            (4, 18.61)
            (5, 18.1)
            (6, 18.79)
            (7, 14.88)
            (8, 17.58)
            (9, 16.41)
            (10, 16.22)
        };

         \addplot[smooth, mark=+, mark size = 5pt, c2, draw opacity = 0.7] plot coordinates {
            (1, 85.09)
            (2, 55.61)
            (3, 51.52)
            (4, 37.49)
            (5, 41.84)
            (6, 34.02)
            (7, 42.64)
            (8, 7.71)
            (9, 9.46)
            (10, 9.97)
        };

        \addplot[smooth, mark=x, mark size = 5pt, c3, draw opacity = 0.7] plot coordinates {
            (1, 18.04)
            (2, 22.91)
            (3, 22.93)
            (4, 16.07)
            (5, 21.96)
            (6, 24.79)
            (7, 16.34)
            (8, 20.7)
            (9, 19.91)
            (10, 18.25)
        };

        \addplot[smooth, mark=star, mark size = 5pt, c4, draw opacity = 0.7] plot coordinates {
            (1, 77.82)
            (2, 44.76)
            (3, 44.75)
            (4, 34.35)
            (5, 36.79)
            (6, 27.25)
            (7, 10.11)
            (8, 9.41)
            (9, 9.58)
            (10, 8.9)
        };
        
        \addplot[smooth, mark=square, mark size = 3pt, c5, draw opacity = 0.7] plot coordinates {
            (1, 78.753)
            (2, 47.740)
            (3, 37.553)
            (4, 43.693)
            (5, 25.733)
            (6, 35.140)
            (7, 28.157)
            (8, 22.843)
            (9, 27.273)
            (10, 22.773)
        };

         \addplot[smooth, mark=triangle, mark size = 3pt, c6, draw opacity = 0.7] plot coordinates {
            (1, 94.960)
            (2, 60.74)
            (3, 44.920)
            (4, 45.460)
            (5, 41.250)
            (6, 37.927)
            (7, 35.087)
            (8, 35.323)
            (9, 34.260)
            (10, 39.503)
        };       
     \end{axis}

    \begin{axis}[
        name = ax3,
        at = {(ax1.south west)},
        yshift = -6cm,
        ylabel = \Huge{Length of trajectory},
        tick label style = {font = \huge},
        every axis plot/.append style={line width = 3pt},
        tick align=outside,
        ymin = 0,
        ytick distance = 5,
        xtick distance = 1,
        xtick pos = left,
        ytick pos = left,
        ymajorgrids,
        grid style = dashed,
        width=3in,
        height=2.5in,
    ]
        \addplot[smooth, mark=o, mark size = 3pt, c1, draw opacity = 0.7] plot coordinates {
            (1, 3.59)
            (2, 3.44)
            (3, 3.69)
            (4, 3.26)
            (5, 3.12)
            (6, 2.97)
            (7, 3.24)
            (8, 2.91)
            (9, 2.63)
            (10, 2.89)
        };

         \addplot[smooth, mark=+, mark size = 5pt, c2, draw opacity = 0.7] plot coordinates {
            (1, 25.45)
            (2, 12.91)
            (3, 13.45)
            (4, 9.15)
            (5, 3.39)
            (6, 2.02)
            (7, 2.02)
            (8, 2.14)
            (9, 2.02)
            (10, 2.02)
        };

        \addplot[smooth, mark=x, mark size = 5pt, c3, draw opacity = 0.7] plot coordinates {
            (1, 5.78)
            (2, 5.38)
            (3, 4.49)
            (4, 4.49)
            (5, 3.51)
            (6, 4.16)
            (7, 4.74)
            (8, 3.89)
            (9, 4.9)
            (10, 5.3)
        };

        \addplot[smooth, mark=star, mark size = 5pt, c4, draw opacity = 0.7] plot coordinates {
            (1, 16.78)
            (2, 10.59)
            (3, 8.79)
            (4, 7.54)
            (5, 9.35)
            (6, 6.85)
            (7, 10.14)
            (8, 2.32)
            (9, 2.01)
            (10, 2.11)
        };
        
        \addplot[smooth, mark=square, mark size = 3pt, c5, draw opacity = 0.7] plot coordinates {
            (1, 14.533)
            (2, 7.767)
            (3, 8)
            (4, 6.5)
            (5, 5.617)
            (6, 6.383)
            (7, 5.883)
            (8, 6.183)
            (9, 5.717)
            (10, 5.65)
        };

         \addplot[smooth, mark=triangle, mark size = 3pt, c6, draw opacity = 0.7] plot coordinates {
            (1, 27.767)
            (2, 23.8)
            (3, 21.633)
            (4, 18.35)
            (5, 13.417)
            (6, 13.467)
            (7, 12.85)
            (8, 11.75)
            (9, 10.233)
            (10, 12.367)
        };  
    \end{axis}

    \begin{axis}[
        name = ax4,
        at = {(ax3.south east)},
        xshift = 2cm,
        tick label style = {font = \huge},
        every axis plot/.append style={line width = 3pt},
        tick align=outside,
        ymin = 0,
        ytick distance = 5,
        xtick distance = 1,
        xtick pos = left,
        ytick pos = left,
        ymajorgrids,
        grid style = dashed,
        width=3in,
        height=2.5in,
    ]
        \addplot[smooth, mark=o, mark size = 3pt, c1, draw opacity = 0.7] plot coordinates {
            (1, 5.07)
            (2, 4.29)
            (3, 3.78)
            (4, 4.06)
            (5, 3.89)
            (6, 4.08)
            (7, 3.33)
            (8, 3.75)
            (9, 3.54)
            (10, 3.49)
        };

         \addplot[smooth, mark=+, mark size = 5pt, c2, draw opacity = 0.7] plot coordinates {
            (1, 19.16)
            (2, 12.28)
            (3, 11.36)
            (4, 8.31)
            (5, 9.54)
            (6, 7.4)
            (7, 9.44)
            (8, 1.91)
            (9, 2.02)
            (10, 2.01)
        };

        \addplot[smooth, mark=x, mark size = 5pt, c3, draw opacity = 0.7] plot coordinates {
            (1, 4.25)
            (2, 5.07)
            (3, 4.98)
            (4, 3.9)
            (5, 5.03)
            (6, 5.5)
            (7, 3.72)
            (8, 4.58)
            (9, 4.42)
            (10, 4.04)
        };

        \addplot[smooth, mark=star, mark size = 5pt, c4, draw opacity = 0.7] plot coordinates {
            (1, 16.94)
            (2, 9.7)
            (3, 9.86)
            (4, 7.63)
            (5, 8.1)
            (6, 5.93)
            (7, 2.24)
            (8, 1.99)
            (9, 2)
            (10, 1.94)
        };
        
        \addplot[smooth, mark=square, mark size = 3pt, c5, draw opacity = 0.7] plot coordinates {
            (1, 17.167)
            (2, 10.150)
            (3, 8.250)
            (4, 9.233)
            (5, 5.533)
            (6, 7.483)
            (7, 6.050)
            (8, 5.000)
            (9, 5.833)
            (10, 5.050)
        };

         \addplot[smooth, mark=triangle, mark size = 3pt, c6, draw opacity = 0.7] plot coordinates {
            (1, 20.500)
            (2, 13.183)
            (3, 9.767)
            (4, 9.817)
            (5, 8.800)
            (6, 8.200)
            (7, 7.617)
            (8, 7.617)
            (9, 7.400)
            (10, 8.550)
        };  
    \end{axis}

    \begin{axis}[
        name = ax5,
        at = {(ax3.south west)},
        yshift = -6cm,
        xlabel = \Huge{Window size $W$},
        x label style = {below = 0.5cm},
        ylabel = \Huge{Single-round reward},
        tick label style = {font = \huge},
        every axis plot/.append style={line width = 3pt},
        tick align=outside,
        ymin = 3,
        ytick distance = 0.5,
        xtick distance = 1,
        xtick pos = left,
        ytick pos = left,
        ymajorgrids,
        grid style = dashed,
        width=3in,
        height=2.5in,
    ]
        \addplot[smooth, mark=o, mark size = 3pt, c1, draw opacity = 0.7] plot coordinates {
            (1, 4.02)
            (2, 4.08)
            (3, 3.54)
            (4, 3.34)
            (5, 4.06)
            (6, 3.92)
            (7, 3.52)
            (8, 3.52)
            (9, 3.94)
            (10, 3.47)
        };

         \addplot[smooth, mark=+, mark size = 5pt, c2, draw opacity = 0.7] plot coordinates {
            (1, 4.43)
            (2, 4.4)
            (3, 4.32)
            (4, 4.28)
            (5, 4.75)
            (6, 4.96)
            (7, 4.94)
            (8, 4.87)
            (9, 4.98)
            (10, 4.94)
        };

        \addplot[smooth, mark=x, mark size = 5pt, c3, draw opacity = 0.7] plot coordinates {
            (1, 4.33)
            (2, 4.63)
            (3, 4.65)
            (4, 4.46)
            (5, 4.65)
            (6, 4.62)
            (7, 4.49)
            (8, 4.65)
            (9, 4.56)
            (10, 4.52)
        };

        \addplot[smooth, mark=star, mark size = 5pt, c4, draw opacity = 0.7] plot coordinates {
            (1, 4.43)
            (2, 4.06)
            (3, 3.88)
            (4, 3.95)
            (5, 4.1)
            (6, 4)
            (7, 4.25)
            (8, 4.94)
            (9, 4.98)
            (10, 4.89)
        };
        
        \addplot[smooth, mark=square, mark size = 3pt, c5, draw opacity = 0.7] plot coordinates {
            (1, 4.733)
            (2, 4.647)
            (3, 4.727)
            (4, 4.653)
            (5, 4.590)
            (6, 4.627)
            (7, 4.647)
            (8, 4.583)
            (9, 4.580)
            (10, 4.580)
        };

         \addplot[smooth, mark=triangle, mark size = 3pt, c6, draw opacity = 0.7] plot coordinates {
            (1, 4.583)
            (2, 4.573)
            (3, 4.577)
            (4, 4.607)
            (5, 4.573)
            (6, 4.537)
            (7, 4.480)
            (8, 4.447)
            (9, 4.527)
            (10, 4.513)
        };  
    \end{axis}

    \begin{axis}[
        name = ax6,
        at = {(ax5.south east)},
        xshift = 2cm,
        xlabel = \Huge{Window size $W$},
        x label style = {below = 0.5cm},
        tick label style = {font = \huge},
        every axis plot/.append style={line width = 3pt},
        tick align=outside,
        ymin = 4,
        ytick distance = 0.25,
        xtick distance = 1,
        xtick pos = left,
        ytick pos = left,
        ymajorgrids,
        grid style = dashed,
        width=3in,
        height=2.5in,
    ]
        \addplot[smooth, mark=o, mark size = 3pt, c1, draw opacity = 0.7] plot coordinates {
            (1, 4.69)
            (2, 4.62)
            (3, 4.65)
            (4, 4.58)
            (5, 4.65)
            (6, 4.65)
            (7, 4.47)
            (8, 4.69)
            (9, 4.64)
            (10, 4.65)
        };

         \addplot[smooth, mark=+, mark size = 5pt, c2, draw opacity = 0.7] plot coordinates {
            (1, 4.44)
            (2, 4.53)
            (3, 4.54)
            (4, 4.51)
            (5, 4.38)
            (6, 4.6)
            (7, 4.52)
            (8, 4.03)
            (9, 4.68)
            (10, 4.96)
        };

        \addplot[smooth, mark=x, mark size = 5pt, c3, draw opacity = 0.7] plot coordinates {
            (1, 4.24)
            (2, 4.52)
            (3, 4.6)
            (4, 4.12)
            (5, 4.37)
            (6, 4.51)
            (7, 4.39)
            (8, 4.52)
            (9, 4.5)
            (10, 4.52)
        };

        \addplot[smooth, mark=star, mark size = 5pt, c4, draw opacity = 0.7] plot coordinates {
            (1, 4.59)
            (2, 4.61)
            (3, 4.54)
            (4, 4.5)
            (5, 4.54)
            (6, 4.6)
            (7, 4.51)
            (8, 4.73)
            (9, 4.79)
            (10, 4.59)
        };
        
        \addplot[smooth, mark=square, mark size = 3pt, c5, draw opacity = 0.7] plot coordinates {
            (1, 4.603)
            (2, 4.690)
            (3, 4.533)
            (4, 4.710)
            (5, 4.650)
            (6, 4.690)
            (7, 4.657)
            (8, 4.577)
            (9, 4.673)
            (10, 4.507)
        };

         \addplot[smooth, mark=triangle, mark size = 3pt, c6, draw opacity = 0.7] plot coordinates {
            (1, 4.627)
            (2, 4.607)
            (3, 4.603)
            (4, 4.627)
            (5, 4.683)
            (6, 4.627)
            (7, 4.603)
            (8, 4.640)
            (9, 4.627)
            (10, 4.620)
        };  
    \end{axis}

     \end{tikzpicture}
\caption{Results under different simulated environments.}
\vspace{-10pt}
\label{fig:robust_window}
\end{figure}

To validate that BiLLP can work robustly in different environment settings, we vary the window size $W$ in the exit mechanism and fix the similarity threshold $\beta$ to simulate different effects of filter bubbles on user disengagement. 
The evaluation results are shown in Figure~\ref{fig:robust_window}, where all results are averaged over three random experiments with distinct seeds. We visualize all three metrics and observe that:
\begin{itemize}[leftmargin=*]
    \item As the window size $W$ increases, the performance of the trajectory length and the cumulative reward metrics decrease for all methods. This implies that when users are more susceptible to the influence of filter bubbles, the model faces greater challenges in learning to improve users' long-term engagement.
    \item BiLLP outperforms all baselines in terms of both the trajectory length and the cumulative reward metrics, which demonstrates the robustness of BiLLP in different environments.
    \item BiLLP obtains a relatively higher score in terms of the single-round in different environments, which indicates it successfully captures user interests while avoiding excessive emphasis on immediate responses.
\end{itemize}

\begin{table}[t]
\small
  \centering
  \caption{Average Results of all methods in the two environments (Bold: Best).}
  \vspace{-10pt}
  \label{tab:experiments_base_models}
  \begin{tabular}{lccc}    
    \toprule
    \multirow{2}*{Methods} & \multicolumn{3}{c}{Steam}\\
    \cmidrule(lr{1em}){2-4}
    & Len & $\mathrm{R_{each}}$ & $\mathrm{R_{traj}}$  \\ 
    \midrule
 GPT-4-32k & \textbf{25.400 $\pm$ 2.800} & \textbf{4.635 $\pm$ 0.115} & \textbf{118.235 $\pm$ 15.915} \\
    GPT-3.5-16k  & 15.367 $\pm$ 0.119 & 4.503 $\pm$ 0.069 & 69.193 $\pm$ 1.590     \\
    Llama-2-7B & 13.800 $\pm$ 1.105 & 4.610 $\pm$ 0.065 & 63.767 $\pm$ 6.015 \\
    \midrule
    \multirow{2}*{Methods} & \multicolumn{3}{c}{Amazon} \\
    \cmidrule(lr{1em}){2-4}
    & Len & $\mathrm{R_{each}}$ & $\mathrm{R_{traj}}$  \\ 
    \midrule
    GPT-4-32k & \textbf{12.450 $\pm$ 1.250} & 4.580 $\pm$ 0.070 & \textbf{57.180 $\pm$ 6.570} \\
    GPT-3.5-16k  & 9.413 $\pm$ 0.190 & 4.507 $\pm$ 0.012 & 42.443 $\pm$ 0.817    \\
    Llama-2-7B & 8.100 $\pm$ 1.512 & \textbf{4.603 $\pm$ 0.054} & 37.300 $\pm$ 6.895 \\
    \bottomrule
  \end{tabular}
  \vspace{-10pt}
\end{table}

\subsubsection{Results with Different Base Models}
To validate the robustness of the BiLLP framework across various base models, we conduct additional experiments with other different LLM backbones: ``gpt-4-32k'' and ``Llama-2-7b''. The results are presented in Table~\ref{tab:experiments_base_models}. From the table, several noteworthy observations can be made:

\begin{itemize}[leftmargin=*]
    \item 
    BiLLP showcases superior performance compared to traditional RL-based methods with different base models, as demonstrated in Table~\ref{tab:experiments_base_models}. This indicates that our framework is robust across different LLMs.
    \item 
    The performance of BiLLP based on ``GPT-3.5-16k'' is superior to that based on ``Llama-2-7B'', while inferior to that based on ``GPT-4-32k''. This observation suggests a positive correlation between the strength of the LLM backbone and the performance enhancement of BiLLP.
\end{itemize}


%

\section{Conclusion}
In this work, we explore the integration of planning capabilities from Large Language Models (LLMs) into the recommendation to optimize long-term engagement. To bridge the gap between the pre-training scenarios and recommendation scenarios, we propose a bi-level learnable LLM planning framework called BiLLP, where the learning process can be divided into macro-learning and micro-learning using a hierarchical mechanism.
This hierarchical approach improves learning efficiency and adaptability. Extensive experiments validate the capability of LLMs to plan for long-term recommendation and the superiority of the BiLLP framework.

A potential avenue for future research involves exploring techniques to enhance the planning capabilities of small-scale models in the context of recommendation tasks. Additionally, exploring the integration of reinforcement learning algorithms within the planning framework could provide further insights into optimizing long-term engagement in recommendation systems.

\begin{acks}
This work is supported by the National Key Research and Development Program of China (2022YFB3104701), the National Natural Science Foundation of China (62272437, 62121002), and the CCCD Key Lab of Ministry of Culture and Tourism.
\end{acks}

\bibliographystyle{ACM-Reference-Format}
\bibliography{reference}

\appendix

\end{document}